# The effect of Si addition on the microstructure and tensile properties of casting Al-5.0Cu-0.6Mn-1.2Fe alloys

Weiwen Zhang, Yuliang Zhao, Datong Zhang, Zongqiang Luo, Chao Yang, Yuanyuan Li

(National Engineering Research Center of Near-net-shape Forming for Metallic Materials, South China University of Technology, Guangzhou 510640, China)

*Corresponding author. Tel: +86-20-87112022, Fax: +86-20-87112111.

E-mail: mewzhang@scut.edu.cn

**Abstract**

In this work, we studied the effect of Si on the microstructure and tensile properties of the as-cast Al-5.0Cu-0.6Mn-1.2Fe alloys which were produced by casting with or without applied pressure. The results show that the addition of Si can significantly influence the microstructure and tensile properties of the alloys. For the alloys produced without pressure, the addition of Si can promote the formation of Chinese script $\alpha$-Fe, suppress the precipitation of plate-like $Al_3(FeMn)$ and Chinese script $Al_6(FeMn)$ and increase the volume percent of porosity, resulting in a remarkable decrease in the ultimate tensile strength (UTS) and yield strength (YS). For the alloys produced with a pressure of 75 MPa, the addition of Si can also promote the formation of fine Chinese script $\alpha$-Fe and high number density $Al_2Cu$ ($\theta$) phases, resulting in a slight increase in UTS and YS. The strength and elongation of the alloys increases with increasing applied pressure at the same Si level, which are attributed to the elimination of porosity, grain refinement strengthening and solid-solution strengthening. The alloy with addition of 1.1 % Si produced under the applied pressure of 75 MPa shows the best tensile properties, where the UTS, YS and elongation is 237 MPa, 140 MPa and 9.8%, respectively.

*Keywords: Al-Cu alloys; Si addition; Tensile properties; Microstructure.*

## 1. Introduction

Casting aluminum alloys, due to their low density and excellent mechanical properties, have been widely used in the transportation and aerospace area, particularly for production of automotive components [1-3]. In addition, recycled Al alloys have received considerable attention since require only 3% of the energy compared to the primary production from natural resources [4]. However, one of the greatest challenges to Al recycling is the high impurity elements in the scraps [5]. Fe is one of the most common impurities in commercial Al alloys which come from bauxite, Fe-based tools and equipment during the production process. As a result of recycling, the Fe levels in these alloys can be invariably increased with the accumulation. However, in the Al-Cu alloys, extremely low Fe content is required, for example less than 0.10 wt.% Fe in A206 series alloys [6]. The solid-state solubility of Fe in $\alpha$-Al is about 0.05 wt.% at equilibrium state [7] and it is even less under real solidification conditions. Consequently, Fe in Al-Cu alloys usually precipitates in the form of the Fe-rich intermetallics, including $Al_3(FeMn)$ [8, 9], $Al_6(FeMn)$ [10, 11], $Al_m(FeMn)$ [8, 11], $Al_7Cu_2Fe$ ($\beta$-Fe) [12-14] and $Al_{15}(FeMn)_3(SiCu)_2$ ($\alpha$-Fe) [15, 16]. Among these Fe-rich intermetallics, the plate-like $Al_3(FeMn)$ and $\beta$-Fe could severely deteriorate the tensile properties of the alloys due to their brittle nature which could result in the crack initiation and propagation [17]. Compared with the coarse plate-like $Al_3(FeMn)$ and $\beta$-Fe, the Fe-rich intermetallic in the form of compact Chinese script has been found to be less harmful to tensile properties [18]. Mn is the most commonly used alloying addition to neutralize the effect of Fe and modify the plate-like Fe-rich intermetallics into less harmful morphologies [19].

Si, like Fe, is another major impurity in many high strength commercial Al-Cu alloys [1]. Recently, interesting research findings [1, 15, 20-23] have been reported concerning Si transfer the Fe-rich intermetallics from platelet-like to less harmful Chinese script. Liu et al. [8, 20, 21] proposed that the addition of Si is helpful to promote the formation of α-Fe and hinder the precipitation of β-Fe because the formation temperature of α-Fe is higher than β-Fe and there are less free Fe atoms available for the formation of β-Fe. Kamguo et al. [22, 23] further confirmed this result. However, Talamantes-Silva et al. [24] reported that β-Fe is the only Fe-rich intermetallic in the 206 alloys with Si/Fe ratio of 1 and Mn/Fe ratio of 3.5. Lemieux et al. [25] and Han et al. [26] observed that the addition of Si does not have an effect on the strength, but loss the ductility of the alloys. However, there are controversial results on the effect of Si on the Al-Cu alloys, i.e. whether Si is an impurity or a beneficial alloying element, or in which range that Si can improve the mechanical properties in Al-Cu alloys. Therefore, it is vital to study the effect of Si on the microstructure evolution and tensile properties of recycled Al-Cu alloys with high Fe and Si contents.

Squeeze casting is a technology with short route, high efficiency and precise forming, possessing features of casting and plastic processing [13, 27, 28], which offers advantages such as enhancing the cooling rate and eliminating casting defects of porosity and shrinkage. Recently, Li et al. [28] reported that squeeze casting can greatly improve the performance of light alloys and reduce the impurity restriction in recycling material. In our previous studies [13, 16], we reported that squeeze casting decreases the secondary dendritic arm spacing (SDAS) of α-Al and the porosity and significantly affects the intermetallics, such as Fe-rich intermetallics, in Al-5.0Cu-0.6Mn alloys.

With accumulation of recycled Al-Cu alloys scraps, the content of Fe and Si contents is relatively high. However, previous studies mainly focused on the effect of low content of Fe and Si contents on the Al-Cu alloys. In present study, the authors comparatively studied the microstructure and tensile properties of Al-Cu alloys with high Fe and Si during squeeze casting produced by casting with or without pressure.

## 2. Experimental procedures

The alloys with various Si contents were produced by melting commercially pure Al (99.5%) and the master alloys of Al-50% Cu, Al-10% Mn, Al-20% Si and Al-5% Fe. The real chemical composition was analyzed by optical emission spectrometer, as shown in Table 1. Firstly, 10 Kg raw materials were melted at about 730 °C in a clay-graphite crucible using an electric resistance furnace and the melts were degassed by 0.5% $C_2Cl_6$ to minimize hydrogen content. The pouring temperature was set at 710 °C after degassing and the die was preheated to approximately 200 °C before squeeze casting. After the melt was poured into a cylindrical die, varied pressure (0, 25 and 75 MPa) was applied to melts and held for 30 s until the melt was completely solidified. Finally, the ingots with a size of Φ 65 mm × 68 mm were obtained.

The tensile samples with dimension of Φ 10 mm × 65 mm were cut from the edge of the ingots. Tensile tests were carried out at room temperature using a SANS CMT5105 standard testing machine with a strain rate of 1 mm/min. At least three samples were tested to obtain the average value. Samples for metallographic observation were cut from the end of tensile specimens. Metallographic samples were etched with 0.5% HF solution for 30 seconds. Samples for grain size measurement were examined in Leica optical microscope with polarized light after anodizing with a 4% $HBF_4$ solution for about 30 s at 20 V. The microstructure was analyzed with the image analysis software Leica Materials workstation V3.6.1. The area fractions of intermetallics were calculated based on the assumption that the morphology of the intermetallics is equaled. Nearly 50 different fields were examined for each sample. The average chemical compositions of the phases and fracture surfaces of tensile specimens were analyzed using SEM Quanta 200, energy-dispersive X-ray analyzer (EDX) and EPMA-1600 electron probe micro-analyzer. X-ray diffraction (Bruker D8 ADVANCE) was used to further index these phases, which was operated at 40 kV and 100 mA. The morphology of the Fe-rich intermetallics was further studied by transmission electron microscopy (TEM) using a Philips microscope (FEI 3010). Cu content in the α(Al) matrix in the as-cast state was measured by EPMA-1600 electron probe micro-analyzer.

## 3. Results and discussion

### 3.1 Microstructure

The grain size distributions of the 0% Si alloys with and without applied pressure after anodizing are displayed in Fig. 1. The red line marked in Fig. 1 is in order to distinguish different grains in the grain boundary. It can be seen that the grain size of the alloy produced with 75 MPa applied pressure is smaller than the alloy produced without applied pressure. The statistics results (Fig. 1c) shows that the increase of applied pressure resulting in a grain size reduction. The measured SDAS of α-Al decreased from ~ 80 μm for the alloy without applied pressure to ~ 30 μm for the alloy with 75 MPa applied pressure. The reduction in grain size is mainly attributed to the applied pressure which can reduce the air gap between the melt and die and an associated increase the cooling rate, leading to local high undercooling and the nucleation rate increasing of primary α-Al [27].

Fig. 2 shows the corresponding microstructure of Al-5.0Cu-0.6Mn-1.2Fe alloys with Si content varying from 0% to 1.1% without applied pressure. As can be seen, the microstructure consists of the primary α-Al, Al$_2$Cu, platelet-like Al$_3$(FeMn), Chinese script Al$_6$(FeMn) and α-Fe, which are confirmed by the EDS results (Table 2). The irregular-shaped light grey phase observed in the microstructure is Al$_2$Cu, which is obtained from the eutectic reaction during solidification. These Fe-rich intermetallics are typical phases found in Al-Cu alloys [8, 13, 30]. As seen from Fig. 2a, the platelet-like Al$_3$(FeMn) and Chinese script Al$_6$(FeMn) precipitate are the main Fe-rich intermetallics in the alloys without Si addition. There is only a small morphology change when the Si content is raised from 0% to 0.15% (Fig. 2b). In the alloy with 0.55% Si, Chinese script α-Fe and Al$_6$(FeMn) coexist (Fig. 2c), while only Chinese script α-Fe is presented when Si content is further increased to 1.1% (Fig. 2d). These indicate that Si promotes the formation of α-Fe and suppresses the formation of $Al_3(FeMn)$ and $Al_6(FeMn)$ phases. Compared with the alloys without applied pressure, the applied pressure influences both the size and morphology of the α-Al and second intermetallics. As shown in Fig. 3, the primary α-Al is effectively refined and the morphology of Fe-rich intermetallics is smaller and more compact in the sample with applied pressure of 75 MPa. As shown in the Fig. 4, the volume percent of porosities in the 1.1% Si alloys is relatively higher than those of alloys in the 0% Si alloy. Moreover, the size of the Fe-rich intermetallics without applied pressure is obviously coarse than those of with applied pressure.

The DSC heating curves of the Al–5.0Cu–0.6Mn-1.2Fe alloys with different Si content are presented in Fig. 5. The DSC curve of the alloy with 0% Si exhibits three distinct endothermic peaks at 656.9, 603.0, and 551 °C, corresponding to the melting of the primary α-Al phases, Chinese script $Al_6(FeMn)$ phase, and eutectic $Al_2Cu$ phases, respectively. Similarly, The DSC curve of the alloy with 1.1% Si exhibits three distinct endothermic peaks at 649.6, 535.2, and 525.5 °C, corresponding to the melting of the primary α-Al phases, eutectic $Al_2Cu$ phases, and eutectic Si phases, respectively. With the increasing Si content, the temperature of liquid and formation of phase are reduced. These results are agreement to the Refs. [20, 23], which indicated that lowered the precipitation temperature of the blocky $Al_2Cu$ phase. The quantitative analysis of volume percent of second intermetallics and porosity as a function of Si content are shown in Fig. 6. These results indicate that the volume percent of Fe-rich intermetallics depends on the Si contents and applied pressure. The volume percent of the total second intermetallics, Chinese Fe-rich intermetallics and porosity increases with increasing Si content, while the volume fraction of platelet-like Fe-rich intermetallics and $Al_2Cu$ decreases with increasing Si content.

Fig. 7 shows the TEM bright-field images of the morphology and distribution of the θ phases in the alloys with different Si addition prepared with applied pressure of 75 MPa. The coarse and needle-like θ phases (~ 800 nm) are uniformly distributed in the α-Al matrix without Si addition (Fig. 7a). With 1.1% Si addition, the size of θ phase is remarkably refined to ~ 100 nm (Fig. 7b) and its EDX result is shown in the Fig. 7c. This indicates that the Si addition promotes the precipitation of high number density θ phase in the alloys and reduces their size. The similar phenomenon was also detected in the Al-Sc-Si [31] and Al-Cu-Si [40] alloys. They considered that the addition of Si could promote the heterogeneous nucleation of the precipitates. The main reason is Si could retard the coarsening of θ precipitates and catalyze the heterogeneous nucleation of θ. Because the Si addition enhance the solubility of Cu in Al, which means less undercooling and lower driving force for the precipitation [40]. Another, Cu could retard the nucleation of Si in the Al-Cu-Si alloys. Once the Si precipitates are formed, they catalyze the nucleation of θ. Si precipitates provide a higher density distribution of heterogeneous nucleation sites and resulting in the precipitation of smaller size θ phase [40]. The other reason cannot be overlook is Si substitutes the place of Cu in the α-Fe ($Al_{15}(FeMn)_3(SiCu)_2$), meaning that there is more Cu atom available for formation of θ phase.

Fig. 8 represents the TEM microstructure and their selected area diffraction patterns (SADP) of the alloys with different Si contents and applied pressure. Fig. 8a show\s that the distribution of platelet-like $Al_3(FeMn)$ with wide of 2 μm adjacent to $Al_2Cu$ in the matrix. The $Al_3(FeMn)$ has a monoclinic structure with a lattice constant of $a = 1.549$ nm, $b = 0.808$ nm, $c = 1.248$ nm. These features are in accordance with the previous works [8, 18]. The Chinese script phase is identified as $Al_6(FeMn)$ in Fig. 8b, which has an orthorhombic structure and lattice parameter of $a = 0.643$ nm, $b = 0.746$ nm, $c = 0.878$ nm. Fig. 8c manifest the Chinese morphology of α-Fe in alloys produced with 0 MPa applied pressure at 1.1% Si. From the SAED pattern analysis, the α-Fe phase, with a body-centered cubic structure, is found to have the lattice parameter of $a = b = c = 1.265$ nm. This phase is also observed in the Al-Si alloy [32].

Fig. 9 shows the samples after deep-etched, revealing the three dimensional morphology of Fe-rich intermetallics. The typical morphology of the deep-etched $Al_3(FeMn)$ phase is long platelet with some branches (Fig. 9a). The Chinese script $Al_6(FeMn)$ phase shows a three dimensional compact skeletal structure (Fig. 9b). Fig. 9c shows the typical eutectic structures of α-Fe with fishbone shape, which is consistent with the previous work [7]. The primary branches grow from center to the side and the secondary branches can grow directly from the sides of the large primary axes. SEM-EDS mapping shown in Fig. 10 reveals that the Si is not only existed in the α-Fe phase but also in α-Al matrix.

The promotion effect of Si on the formation of α-Fe can be summarized as follows. Firstly, Si could react with Mn and Fe to form α-Fe when the enough amount of these alloying elements existing in the alloy [20, 33]. Secondly, the grow rate of monoclinic structure of $Al_3(FeMn)$ and orthorhombic structure $Al_6(FeMn)$ is slower than the body-centered cubic structure α-Fe [8]. Since the presence of Si have a strong restriction factor [34], which is helpful to the formation of compact and cubic α-Fe and suppress the formation of two-dimensional plate-like $Al_3(FeMn)$ and, α-Fe can nucleate and grow quickly on the three-dimensional network of dendrite arm (Fig. 8c) in any direction [8]. Thirdly, increasing Si content reduces the liquidus temperature (Fig. 4), resulting in the easy to formation of α-Fe, owing to the formation temperature of $Al_3(FeMn)$ and $Al_6(FeMn)$ is higher than α-Fe [9]. According to the ref [8, 20], the formation temperature of $Al_3(FeMn)$ and $Al_6(FeMn)$ is both above 640 °C, but the liquidus temperature of the alloy with 1.1% Si is 649.6 °C. Because of the relatively high cooling rate during the die casting, the formation of the $Al_3(FeMn)$ and $Al_6(FeMn)$ can be suppressed. Thus, there are more free available Fe atom for the formation of α-Fe.

### 3.2 Tensile properties

Fig. 11 shows the effect of applied pressure and Si content on the tensile properties of the alloys. It is worth emphasizing that an increase in applied pressure brings a significant improvement in the strength and elongation of Al-5.0Cu-0.6Mn-1.2Fe alloys. For the alloys without the addition of Si, when the applied pressure increase from 0 MPa to 75 MPa, the UTS of the alloys increase from 206 MPa to 223 MPa, the YS increase from 109 MPa to 114 MPa and the elongation increase from 6.1 % to 11.7 %, respectively. This result is consistent with previous study [35]. The increments of UTS, YS and elongation of the alloys are about 8.1%, 4.6%, and 93.1%, respectively. Similarly, for the alloys with addition of 1.1% Si, when the applied pressure increase from 0 MPa to 75 MPa, the increments of UTS, YS and elongation are further increased to 35.2%, 40.0%, and 140.2%, respectively.

The fracture surface of the alloys with different Si contents is shown in Fig. 12. The fracture surfaces of the Si-free alloys without applied pressure (Fig. 12a and b) show many large porosities (~ 50 µm) and high volume fraction of second phases at grain boundaries, leading to many tearing ridges. As the applied pressure increases to 75 MPa, some dimples but no porosity are found in the fracture surface (Fig. 12c and d), which implies that the applied pressure tend to improve the ductility. The fracture surface of the alloy with 1.1% Si produced without applied pressure shows the porosities and microcracks and the cracks propagate along the interface between large α-Fe and α-Al (Fig. 12e and f). For the samples with applied pressure of 75 MPa, the fracture morphology of integrated α-Fe and $Al_2Cu$ phases shows the feature of quasi-cleavages fracture, as seen in Fig. 11g and h. In order to further understand the failure mechanism, the longitudinal microstructure beneath the fracture surfaces of the alloys is shown in Fig. 13. The second phases and porosities are the preferable location for the crack initiation and propagation (Fig. 13a and c), while cracks mainly concentrated on refined and compact α-Fe for the samples produced by squeeze casting (Fig. 13b and d).

As described above, the strength and elongation of the alloys increases with increasing applied pressure at the same level of Si, which could be deduced from the grain refinement strengthening, solid-solution strengthening and elimination of porosity. First, the average grain size of the alloy under 75 MPa applied pressure decreases comparing with the alloy without applied pressure. Also, the solid-solution strengthening is contributed to the increasing of the strength [36-39]. The solubility of the alloying elements in the α-Al matrix is presented in Table 3 and the microhardness of the solid-solution matrix was measured, as displayed in Fig. 14. Take 1.1% Si alloy for example, as the applied pressure increases from 0 to 75 MPa, the hardness of matrix increases from 64.4 to 75.7 HV and the solubility of Cu in α-Al matrix increase from 1.24 to 2.59 at.%. These implies that the applied pressure makes higher amount of Cu dissolve in to the α-Al matrix leading to the solid-solution strengthening [36]. In addition, the elimination of porosity partially contributes to the improvement of strength.

The UTS, YS and elongation of the alloys decreased with increasing Si content when the alloys were produced without applied pressure. The increasing percent of porosity and Fe-rich intermetallics play a leading role on the deterioration of the tensile properties. As shown in Fig. 4, the porosity area increases as the Si content increases. The cracks are easily initiation and propagate along the porosity, resulting in the decreasing of UTS and YS. The increasing amount of Fe-rich intermetallics prevents liquid feeding and promotes formation of porosity in the last solidification The site of porosity and coarse Fe-rich intermetallics are apt to be the cracks initiation, as demonstrated in the fracture surface (Fig. 12a and e). It is interesting to note that the UTS, YS and elongation of the alloys increased slightly with increasing Si content when the alloys produced at the applied pressure of 75 MPa. It is considered that Si addition promotes the formation of the Chinese script α-Fe and suppresses the formation of plate-like $Al_3(FeMn)$ and Chinese script $Al_6(FeMn)$, which is less harmful to the tensile properties. The Chinese script α-Fe is usually interwoven with α-Al, making the crack more difficult to propagate. These are also confirmed by the cracks propagation through the arms of α-Fe in multi-directions (Fig.11e and g) and the broken Fe-rich intermetallics (Fig.12b and d). And the applied pressure makes the Fe-rich intermetallics tend to be refined and compact, which could strengthen the alloy by acting as pins to prevent dislocations from sliding under stress. Also, Si enhances the strength of the alloys by increasing the precipitate number density of fine θ phase. Si addition increases the precipitate number density of θ phase (Fig. 7) and reduces its size because Si addition could accelerate the precipitation kinetics and coarsening resistance of the precipitate [31].

**5. Conclusions**

The microstructure and tensile properties of the squeeze-cast and gravity cast Al-5.0Cu-0.6Mn-1.2Fe alloys with different Si contents have been studied comparatively. The Al-5.0Cu-0.6Mn-1.2Fe alloys show three kinds of Fe-rich intermetallics, i.e. Chinese script α-Fe, $Al_6(FeMn)$ and platelet-like $Al_3(FeMn)$. Increasing Si content to 0.55% promotes the formation of less harmful α-Fe and suppresses the formation $Al_3(FeMn)$ and $Al_6(FeMn)$. The applied pressure significantly improves the strength of the alloys, which is attributed to the elimination of porosity, grain refinement strengthening and solid-solution strengthening. The UTS and YS of the alloys produced at 75 MPa increases obviously with the increasing Si content, while the UTS and YS decrease without applied pressure, owing to the Si addition promotes the formation of compact α-Fe and the precipitation of refined and high number density θ phases under applied pressure. In addition, the alloy with addition of 1.1 % Si produced under the applied pressure of 75 MPa shows the best tensile properties, where the UTS, YS and elongation is 237 MPa, 140 MPa and 9.8%, respectively, which is 35.2%, 40.0%, and 140.2% higher than those of the counterpart without applied pressure, respectively.

**Acknowledgement**

The financial support from Natural Science Foundation of China (51374110), Major Special Project for Science and Technology of Guangdong Province (No. 2015B090926004) and Team project of Natural Science Foundation of Guangdong Province (2015A030312003) are acknowledged. We are grateful to the Chinese Scholarship Council (CSC) for financial support of this work.

**Figures caption**

**Fig. 1.** Effect of applied pressure on the grain size of the alloy containing 0% Si：(a) without applied pressure; (b) with 75 MPa applied pressure; (c) histogram showing the distribution of grain size of the alloy with and without applied pressure.

**Fig. 2.** Microstructures of the as-cast alloys at applied pressure of 0 MPa: (a) 0% Si; (b) 0.15% Si; (c) 0.55% Si; (d) 1.1% Si.

**Fig. 3.** Microstructure of the as-cast alloys at applied pressure of 75 MPa: (a) 0% Si; (b) 0.15% Si; (c) 0.55% Si; (d) 1.1% Si.

**Fig. 4.** The porosities in the alloys at low magnification: (a) 0% Si; (a) 1.1% Si.

**Fig. 5.** DSC heating curve of the alloys with different Si alloys.

**Fig. 6.** Microstructural features in the alloys with different applied pressures and Si content: (a) total volume percent of second intermetallics; (b) volume percent of needle-like Fe-rich intermentallics; (c) volume percent of Chinese-script Fe-rich intermentallics; (d) volume percent of $\theta$ ($Al_2Cu$) and (e) volume percent of porosity.

**Fig. 7.** TEM micrographs of alloys produced by applied pressure of 75 MPa without and with Si addition: (a) without Si addition; (b) with 1.1 % Si addition; (c) EDS analysis results of $Al_2Cu$.

**Fig. 8.** Microstructure of the experimental samples examined by TEM: (a) 0% Si 0 MPa ($Al_3(FeMn)$); (b) 0% Si 75 MPa ($Al_6(FeMn)$); (c) 1.1% Si 0 MPa (α-Fe); (d) 1.1% Si 75 MPa (the dislocation around α-Fe).

**Fig. 9.** Typical three dimensional morphology of the Fe-rich intermetallics: (a) platelet $Al_3(FeMn)$; (b) Chinese script $Al_6(FeMn)$; (c) Chinese script α-Fe.

**Fig. 10.** The SEM mapping of α-Fe phase.

**Fig. 11.** Mechanical properties of alloys at different Si content and applied pressures: (a) UTS; (b) YS; (c) Elongation.

**Fig. 12.** Fracture surfaces of the alloys with different Si contents and applied pressures: (a, b) 0% Si 0 MPa; (c, d) 0% Si 75 MPa; (e, f) 1.1% Si 0 MPa; (g, h) 1.1% Si 75 MPa.

**Fig. 13.** The Profiles beneath the fracture surfaces of the alloys at different applied pressures and silicon contents: (a) 0% Si 0 MPa; (b) 0% Si 75 MP; (c) 1.1% Si 0 MPa; (d) 1.1% Si 75 MPa.

**Fig. 14.** The microhardness of the α-Al matrix in differnt alloys.

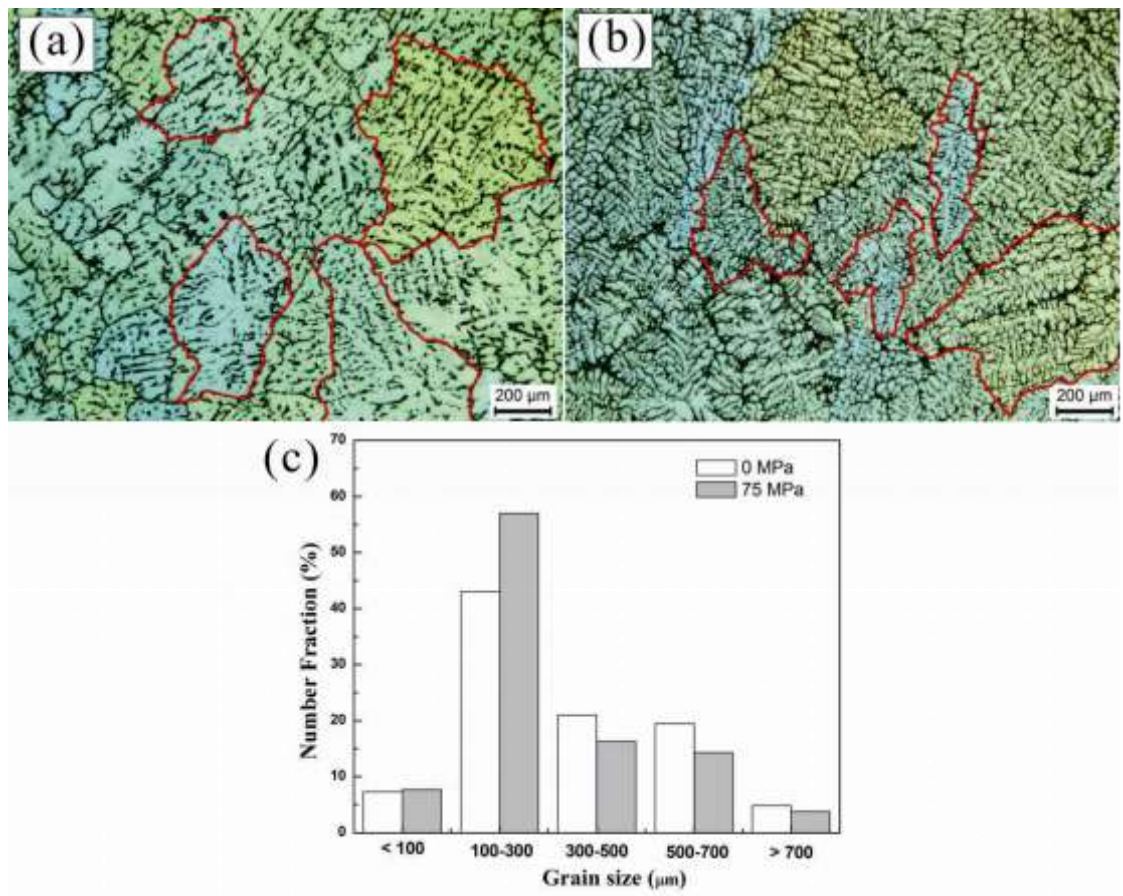

**Fig. 1.**

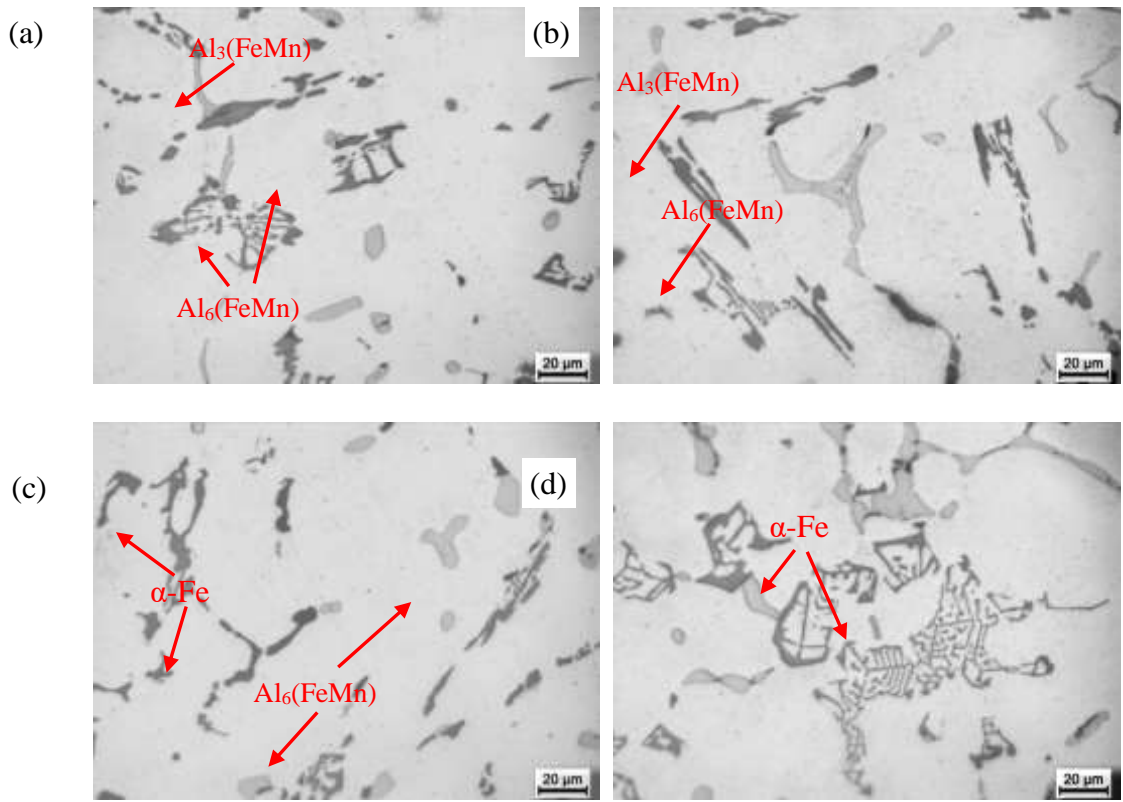

**Fig. 2.**

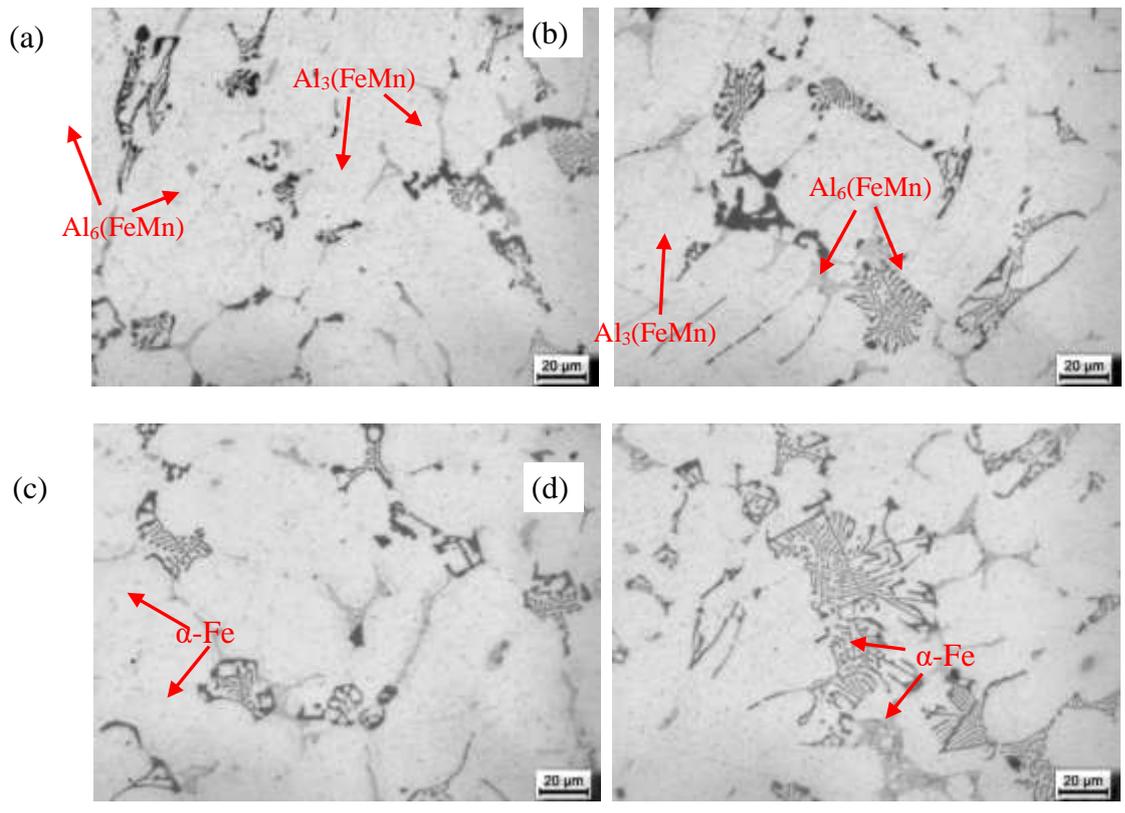

**Fig .3.**

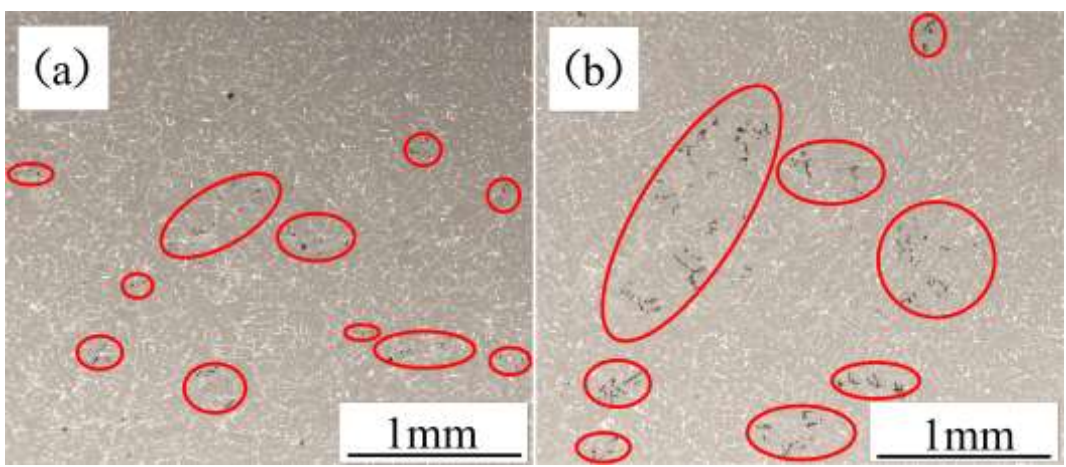

**Fig. 4.**

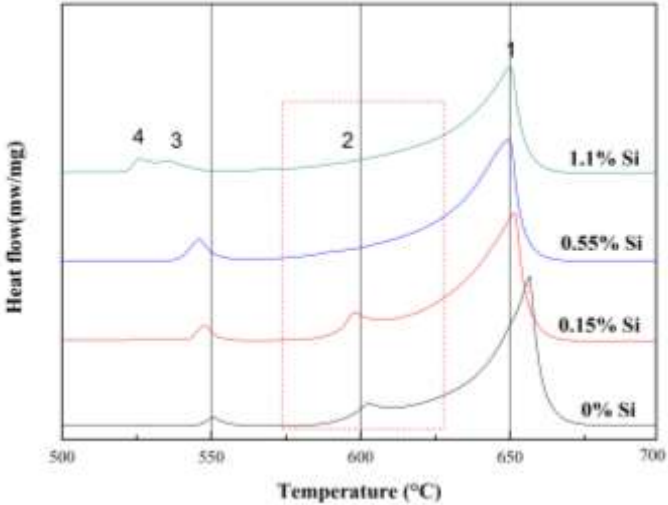

**Fig. 5.**

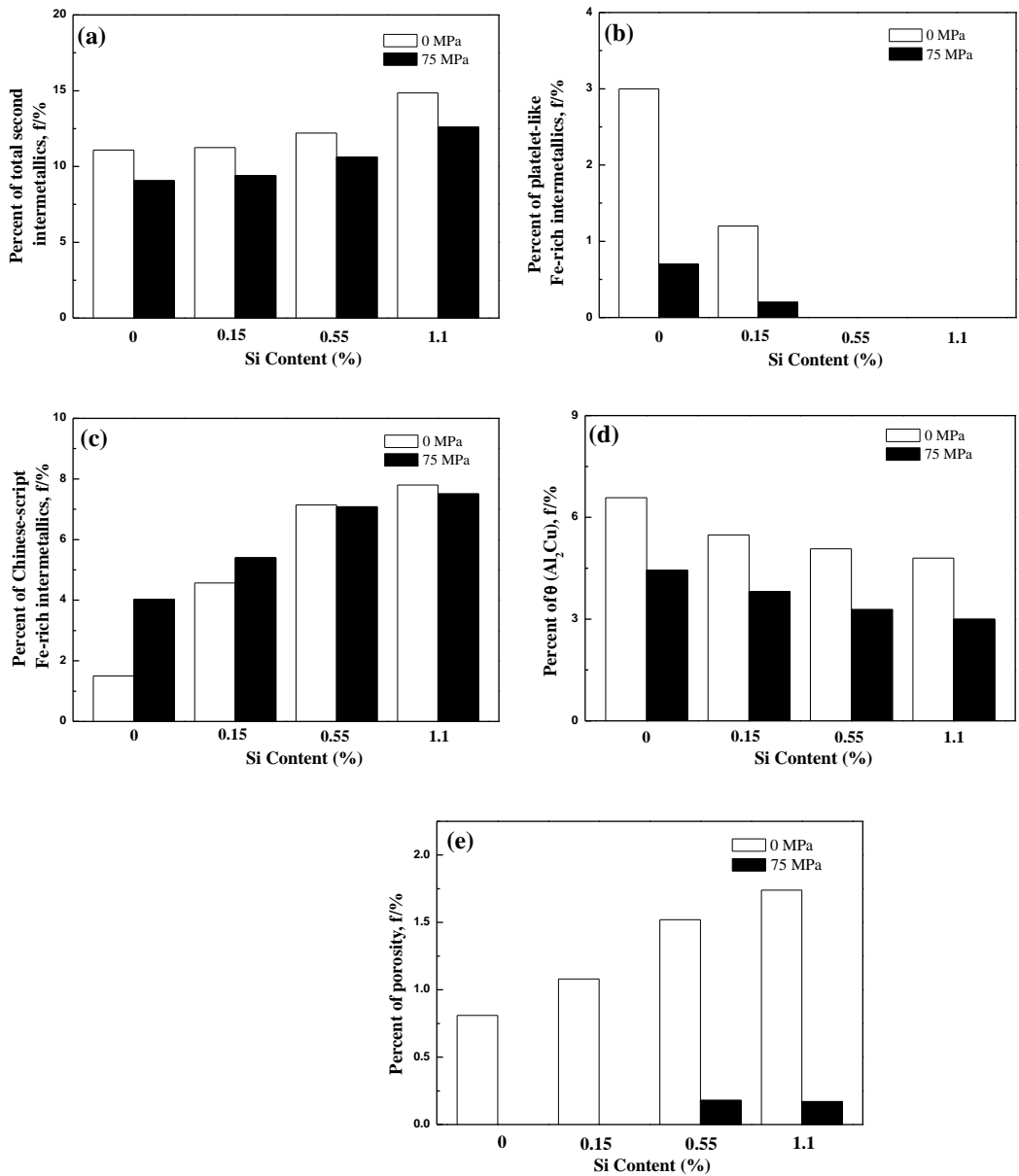

Fig. 6.

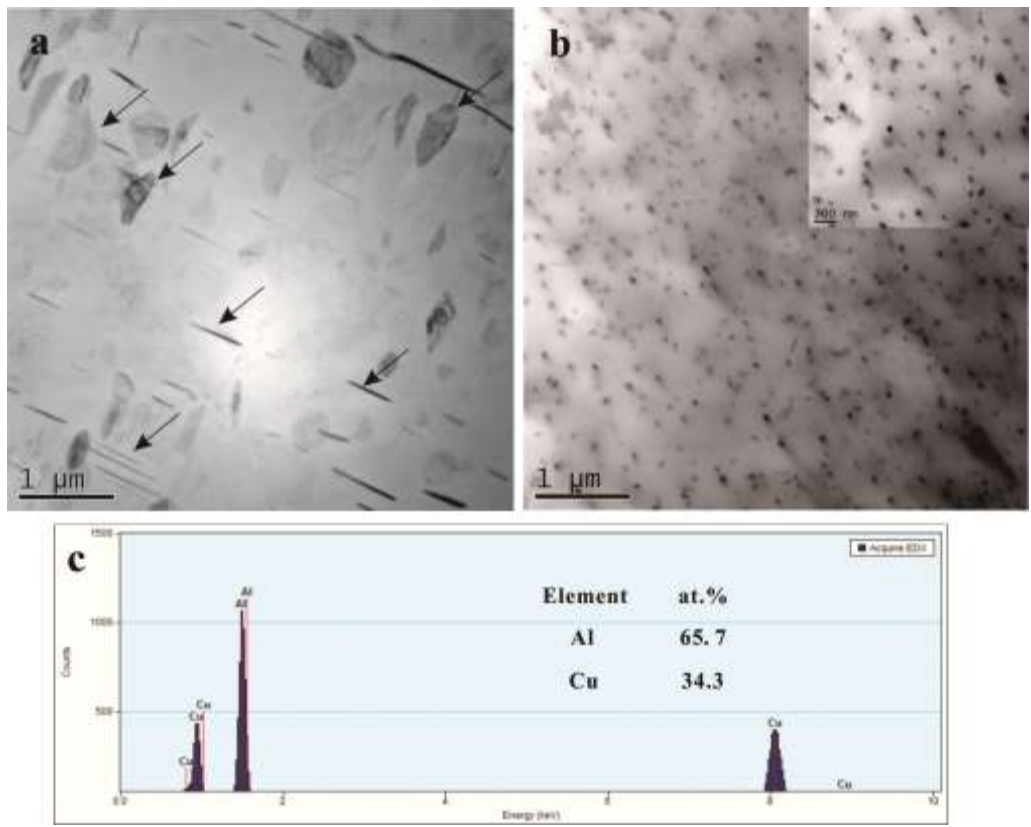

**Fig. 7.**

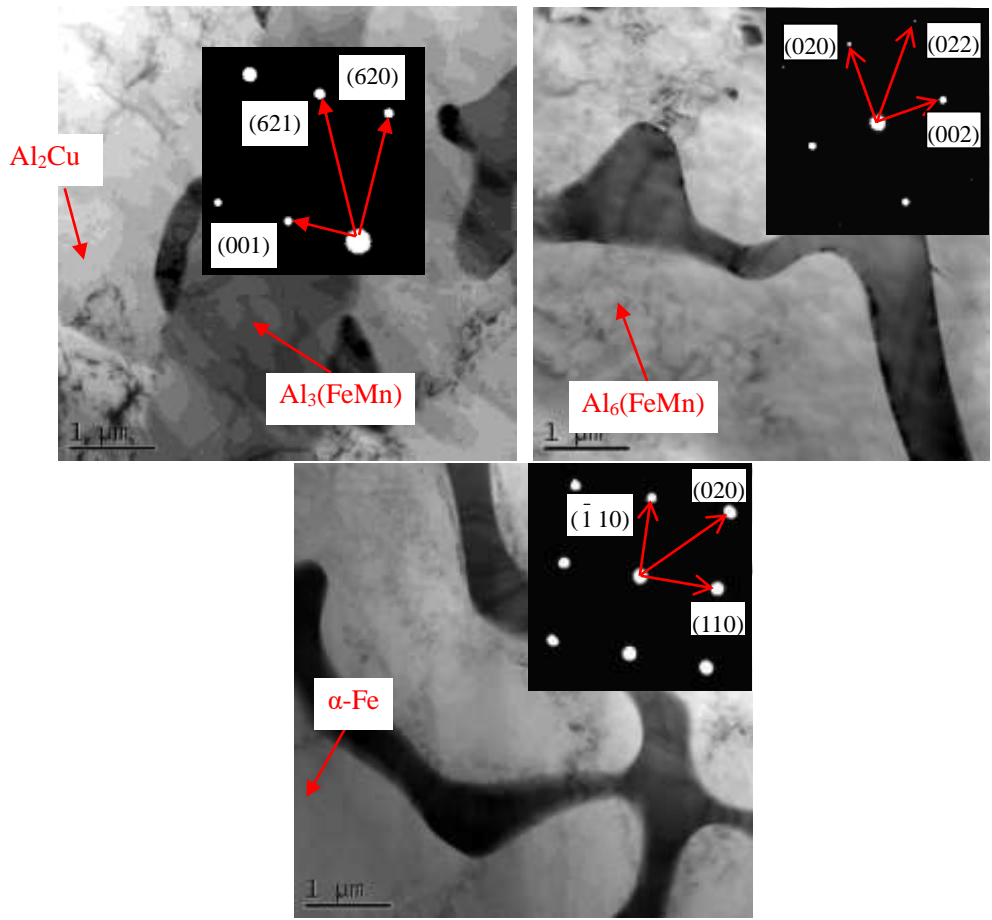

**Fig. 8.**

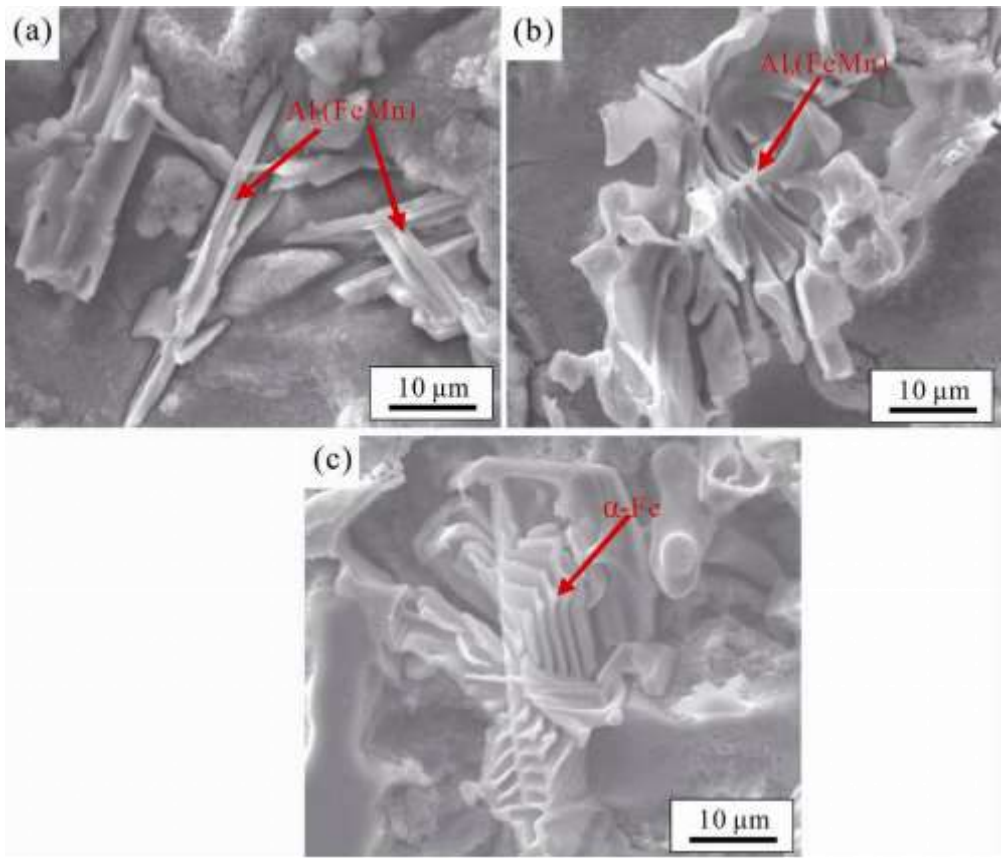

**Fig. 9.**

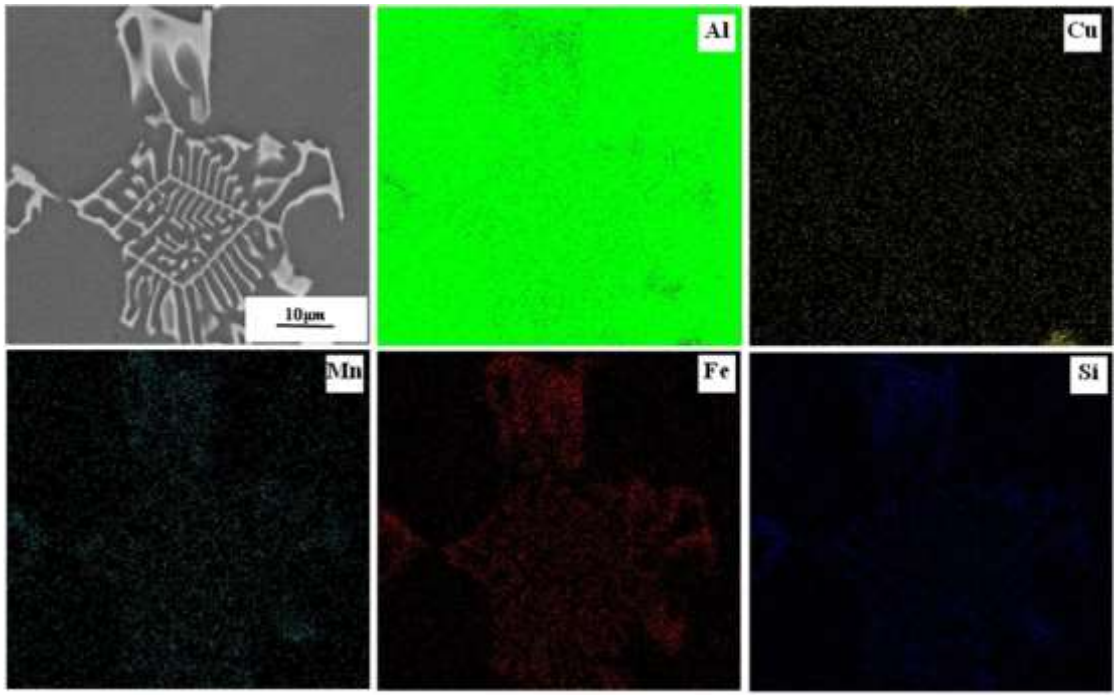

**Fig. 10.**

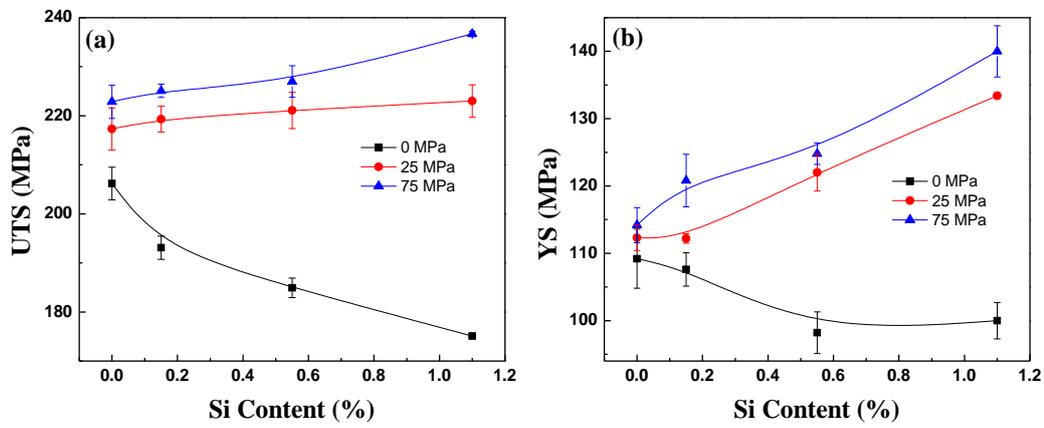

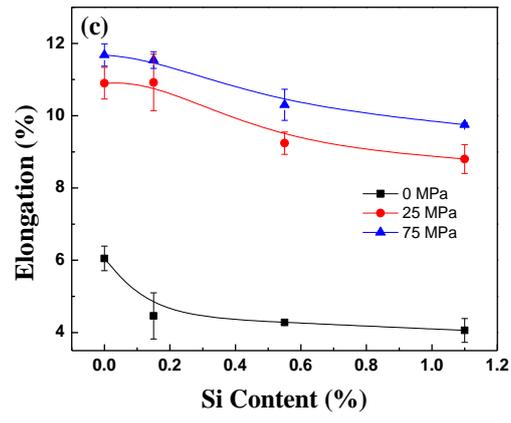

Fig. 11.

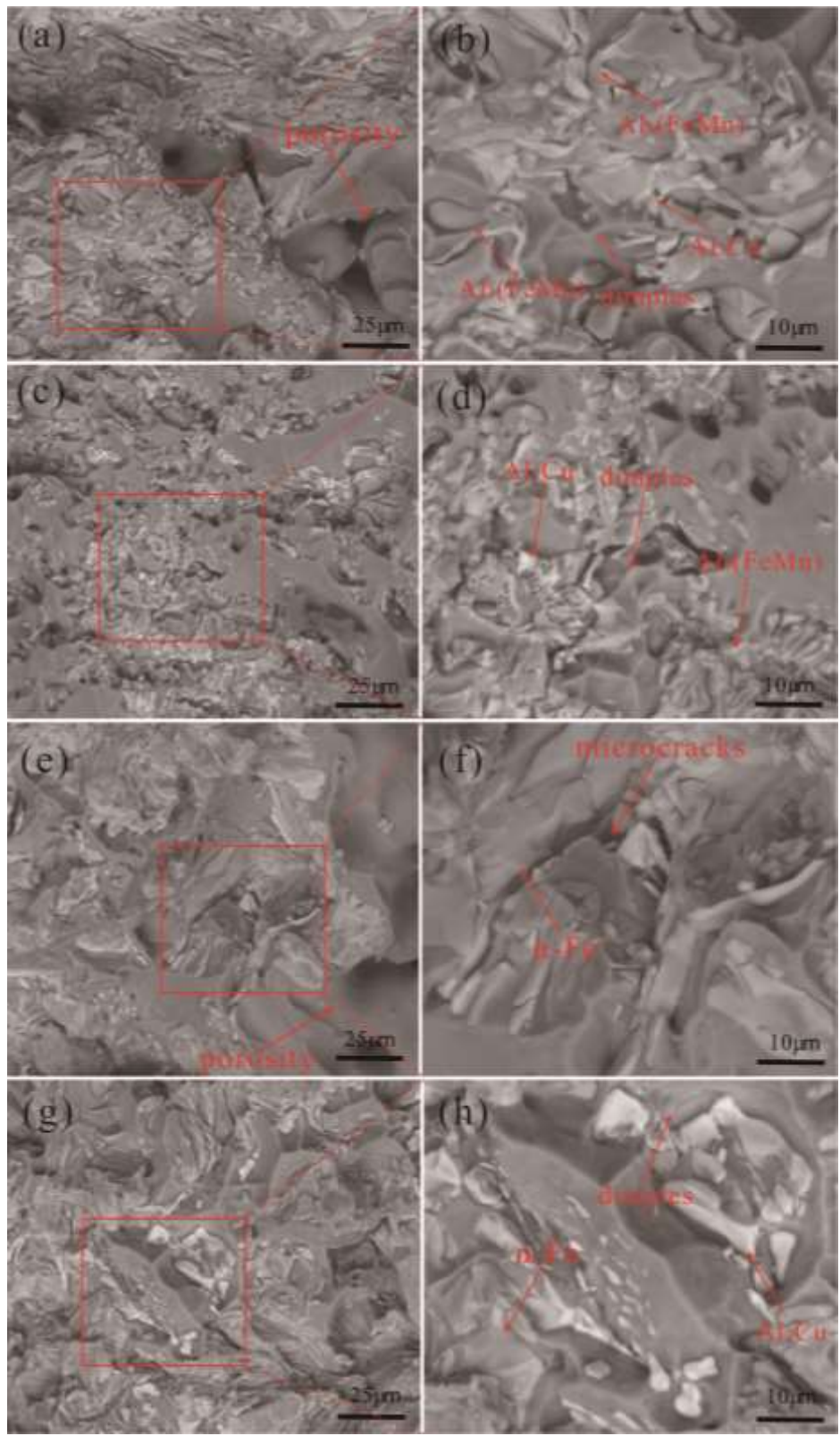

**Fig. 12.**

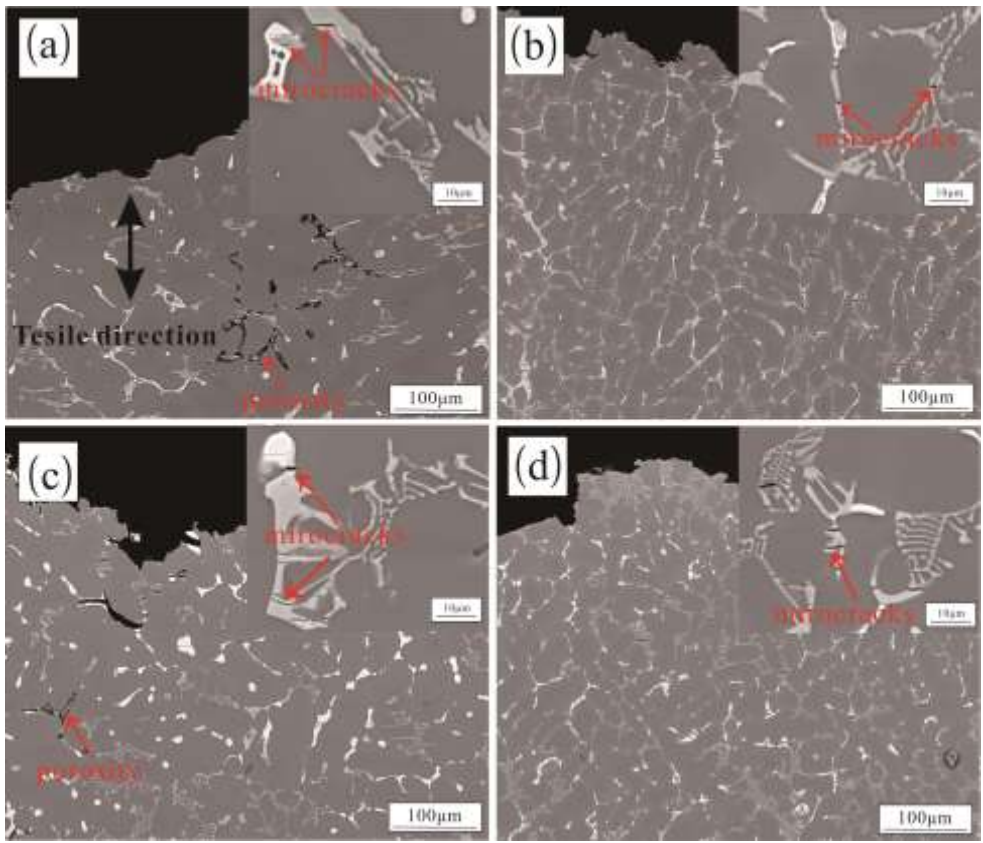

**Fig. 13.**

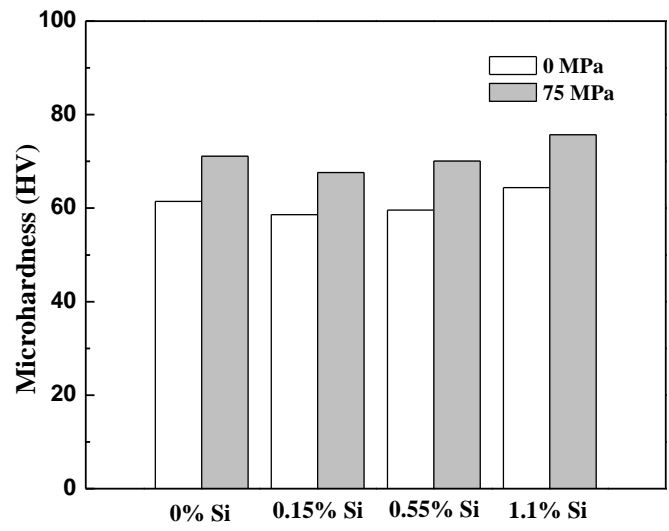

**Fig. 14.**